\documentclass[aps,pre,preprint,superscriptaddress,showpacs]{revtex4}
\usepackage{graphicx}
\usepackage{amsmath}
\usepackage{amssymb}
\usepackage{dsfont}
\usepackage{subfigure}
\usepackage{mathptmx}
\usepackage{color}

\newcommand{\altered}{}
\newcommand{\alteredRWN}{}
\begin{document}

\renewcommand{\vec}[1]{{\bf #1}}
\newcommand{\eye}{\mathds{1}}

\title{Singular forces and point-like colloids in lattice Boltzmann
hydrodynamics}

\author{R. W. Nash}
\affiliation{SUPA, School of Physics, The University of Edinburgh,
  JCMB King's Buildings, Edinburgh EH9 3JZ, United Kingdom}
\author{R. Adhikari}
\affiliation{The Institute of Mathematical Sciences, CIT Campus,
  Tharamani, Chennai 600113, India}  
\affiliation{SUPA, School of Physics, The University of Edinburgh,
  JCMB King's Buildings, Edinburgh EH9 3JZ, United Kingdom} 
\author{M. E. Cates}
\affiliation{SUPA, School of Physics, The University of Edinburgh,
  JCMB King's Buildings, Edinburgh EH9 3JZ, United Kingdom}

\begin{abstract} We present a second-order
accurate method to include arbitrary distributions of force densities
in the lattice Boltzmann formulation of hydrodynamics. Our method may
be used to represent singular force densities arising either from
momentum-conserving internal forces or from external forces which do
not conserve momentum.  We validate our method with several examples
involving point forces and find excellent agreement with analytical
results.  A minimal model for dilute sedimenting particles is
presented using the method which promises a {\altered substantial gain
in computational efficiency}.
\end{abstract}


\maketitle

\section{Introduction} The numerical integration of the discrete velocity 
Boltzmann equation provides an efficient method for the solution of 
isothermal, incompressible fluid flows in complex geometries \cite{Succi:2001}.
The finite-difference equation generated by the integration scheme is referred 
to as the lattice Boltzmann equation (LBE). The method can be extended to 
study multiphase \cite{Shan:1994} and multicomponent \cite{Swift:1996} flows, 
the hydrodynamics of polymers \cite{Ahlrichs:1998} and suspensions 
\cite{Ladd:1994, Ladd:1994b}, and flows under gravity \cite{Buick:2000}.

In the extensions of the LBE described above, the resulting momentum balance
equations contain additional terms beyond the usual pressure and viscous
forces. These represent forces acting on the fluid, either from external
sources like gravity, or internal sources like the gas-liquid interface in a
two-phase fluid. External sources can add to the total momentum of the fluid,
while internal sources, being immersed within the fluid, can only exchange
momentum with it. Internal forces on the fluid, therefore, can always be
expressed as the divergence of a stress tensor. This formulation encodes the
fact that there are no local sources or sinks of momentum.

The LBE, derived as it is from the Boltzmann equation for a dilute
gas, can only faithfully represent the hydrodynamics of a fluid with
an ideal-gas equation of state and a Newtonian constitutive equation
\cite{Resibois:1977}. {\altered One} way around this restrictive
situation is to use the forced Boltzmann equation to represent the
additional forces that appear in the extensions described above
\cite{Succi:2001}. So far, this idea has been used mainly to model the
effects of gravity \cite{Buick:2000} and gas-liquid interfacial forces
in non-ideal gases \cite{Shan:1994}.  These correspond to two special
types of force distributions: in the case of gravity, the force is
spatially and temporally constant, while in the case of the gas-liquid
interface, the force varies both in space and time, but is only
evaluated on the nodes of the computational grid. The forces in either
case are \emph{smooth} functions of position.

However, many models of boundaries immersed in fluids require
\emph{singular} distributions of forces. Such a description follows,
for example, when a gas-liquid interface is described as a
two-dimensional manifold of zero thickness instead of a
three-dimensional volume of space where the density changes
rapidly. In a similar mathematical idealization, a polymer in a fluid
may be represented as a one-dimensional curve with a singular
distribution of forces \cite{Doi:1986}. In yet another example, at
distances large compared to its radius, a sedimenting colloid can be
well-approximated as a singular point force~\cite{Saffman:1973}.
Clearly, the range of applications of lattice Boltzmann hydrodynamics
can be greatly expanded if singular force densities, not necessarily
located at grid points, can be incorporated into the method.

In this paper we show how to include force densities having smooth or
singular distributions, located at arbitrary (in general, off-lattice)
points into the lattice-Boltzmann formulation of hydrodynamics. In the
following section we first discuss the discrete representation of the
forcing term in the Boltzmann equation and derive a second-order
accurate integration scheme for the discrete velocity forced Boltzmann
equation using the method of characteristics. In Section
\ref{sec:singular} we introduce a general distribution of singular
forces and using a {\altered suitable} regularization of the delta
function obtain a smooth but sharply peaked distribution of
forces. The method is exemplified in Section \ref{sec:singular} for
three common singular force distributions (a Stokeslet, a stresslet,
and a rotlet) and validated for the Stokeslet case by comparison with
fully resolved numerical simulation.  Finally, we show how our method
can be adapted to provide a simplified description of a dilute
suspension of sedimenting colloids. We end with a summary of our
method and discuss potential applications.

\section{Multiple relaxation time forced lattice-Boltzmann equation}
\label{sec:MRT}
The LBE may be derived from the Boltzmann equation by a two-step
procedure. First, a discrete velocity Boltzmann equation (DVBE) is
obtained by retaining a finite number of terms in the Hermite
expansion of the Boltzmann equation and evaluating the conserved the
moments using a Gauss quadrature \cite{He:1997}. The discrete
velocities $\{{\bf c}_i\}$ are the nodes of the Gauss-Hermite
quadrature. This is followed by a discretization in space and time to
provide a numerical integration scheme, which is commonly called the
LBE \cite{Chen:1998b}.

Usually a first-order explicit Euler scheme is used to integrate the
DVBE, which surprisingly enough, gives second-order accurate results
\cite{Sterling:1996}. This is so because the discretization error has
the same structure as the viscous term in the Navier-Stokes equation,
whereby it can be absorbed by a simple redefinition of the viscosity
to give second-order accuracy. The same Euler scheme for the forced
Boltzmann equation gives a discretization error term which can be
absorbed only be redefining physical quantities like the momentum and
stress \cite{Ladd:2001}. Below, we provide a a straightforward
explanation of these redefinitions and show how they are related to
the discretization error induced by the integration scheme.

We begin with the discrete velocity Boltzmann equation including an
external acceleration field ${\bf F}({\bf x},t)$
\begin{equation}
\label{dbe}
\partial_t f_i +{\bf c}_i\cdot\nabla f_i  + [{\bf 
F}\cdot\nabla_{\bf c}f]_i = - {\cal L}_{ij}(f_j - f_j^0)
\end{equation}
where $f_i({\bf x},t)$ is the one-particle distribution function in
phase space of coordinates ${\bf x}$ and velocities ${\bf c}_i$,
${\cal L}_{ij}$ is the collision matrix linearized about the local
equilibrium $f^0_i$ and the repeated index $j$ is summed over. Mass
and momentum conservation require the collision term to satisfy
\begin{equation}
\label{conservation}
\sum_{i=0}^{n}{\cal L}_{ij}(f_j - f_j^0) = 0,\qquad
\sum_{i=0}^{n}{\cal L}_{ij}(f_j - f_j^0){\bf c}_i = 0
\end{equation}
while isotropy requires that the ${\cal L}_{ij}$ depend only on the
angles between ${\bf c}_i$ and ${\bf c}_j$
\cite{Succi:2001}. Eq.~(\ref{dbe}) is most easily derived by expanding
the distribution functions in terms of tensor Hermite polynomials,
truncating the expansion at a certain order, and evaluating the
expansion coefficients using a Gaussian quadrature \cite{He:1997e}. In
$d$ dimensions, the quadrature is defined by the $n$ discrete
velocities ${\bf c}_i$ and a set of weights $w_i$ giving rise to a
$DdQn$ discrete Boltzmann equation \cite{Qian:1992}. Retaining terms
up to second order in the Hermite expansion is sufficient for
isothermal fluid flow problems.  The equilibrium distribution
functions to second-order in the Hermite expansions are

\begin{equation}\label{eqf}
f_i^0(\rho,{\bf v}) = w_i\left[\rho + \frac{\rho{\bf v}\cdot{\bf
c}_i}{c_s^2} + \frac{\rho{\bf vv}:{\mathbf Q}_i}{2c_s^4}\right]
\end{equation}

where the tensor $Q_{i\alpha\beta}\equiv
c_{i\alpha}c_{i\beta}-c_s^2\delta_{\alpha\beta}$ (where Greek indices
denote Cartesian directions) and $c_s$ is the speed of sound. The mass
density $\rho$ and the momentum density $\rho\vec{v}$ are moments of
the distribution function:

\begin{equation}
\label{macrovariable1}
\rho = \sum_{i=0}^{n}f_i,\qquad \rho{\bf v} = \sum_{i=0}^{n}f_i{\bf c}_i. 
\end{equation}
To the same order in the Hermite expansion, the discrete
representation of the forcing term is given by~\cite{PhysRevE.58.6855}
\begin{eqnarray}
\nonumber [{\bf F}\cdot\nabla_{\bf c}f]_i &=& - \rho
w_i\left[\frac{{\bf F}\cdot{\bf c}_i}{c_s^2}+\frac{({\bf vF}+{\bf
Fv}):{\mathbf Q}_i}{2c_s^4}\right]\\ &\equiv&-\Phi_i({\bf x},t)
\end{eqnarray}
Finally, the deviatoric momentum flux tensor
\begin{equation}
S_{\alpha\beta} = \Pi_{\alpha\beta}-\rho
c_s^2\delta_{\alpha\beta}=\sum_{i=0}^n f_i Q_{i\alpha\beta}
\end{equation}
 is the second moment of the distribution function. In isothermal
models, the higher moments represent non-conserved kinetic degrees of
freedom, commonly known as ghost modes. In the hydrodynamic limit,
Eq.~(\ref{dbe}) gives rise to Navier-Stokes behaviour, described by
\begin{equation}
\rho(\partial_t {\bf v} + {\bf v}\cdot\nabla{\bf v}) = -\nabla p
+\eta\nabla^2{\bf v} + \zeta\nabla(\nabla\cdot{\bf v}) + {\bf F}
\end{equation}
where the pressure obeys $p = \rho c_s^2$, and the shear viscosity
$\eta$ and the bulk viscosity $\zeta$ are related to the eigenvalues
of ${\cal L}_{ij}$. {\altered In practice the algorithm is normally
used in a parameter regime where the fluid is nearly incompressible
($\nabla\cdot{\bf v} \simeq 0$).}

To begin our derivation of the numerical scheme we rearrange
Eq.~(\ref{dbe}) to obtain
\begin{equation}
\label{dbe1}
\partial_t f_i +{\bf c}_i\cdot\nabla f_i = R_i({\bf x},t)
\end{equation}
where $R_i({\bf x},t) = -{\cal L}_{ij}[f_j({\bf x},t) - f_j^0({\bf
x},t)]+ \Phi_i({\bf x},t)$ represents the effects of both collisions
and forcing. Eq.~(\ref{dbe1}) represents a set of first-order
hyperbolic equations and can be integrated using the method of
characteristics \cite{zwillinger}. Integrating over a time interval
$\Delta t$ we have
\begin{equation}
\label{lbe1}
f_i({\bf x} + {\bf c}_i\Delta t, t+\Delta t) -  f_i({\bf x},t) = 
\int_{0}^{\Delta t}ds R_i({\bf x}+{\bf c}_is, t+s)
\end{equation}
The integral above may be approximated to second-order accuracy using
the trapezium rule and the resulting terms transposed to give a set of
implicit equations for the $f_i$:
\begin{eqnarray}
f_i({\bf x} + {\bf c}_i\Delta t, t+\Delta t) - \frac{\Delta
t}{2}R_i({\bf x}+{\bf c}_i\Delta t, t+\Delta t) = \nonumber \\
f_i({\bf x},t) -\frac{\Delta t}{2}R_i({\bf x}, t) + \Delta t R_i({\bf
x}, t)
\end{eqnarray}
The structure of the above set of equations suggests the introduction
of a new set of \emph{auxiliary} distribution functions
\cite{Shan:1994}, \cite{Dellar:2001}
\begin{eqnarray}
\label{transformf}
\bar f_i({\bf x},t) = f_i({\bf x},t) - \frac{\Delta t}{2}R_i({\bf x},t) 
\end{eqnarray}
in terms of which the previous set of equations are explicit,
\begin{equation}\label{halftransform}
\bar f_i({\bf x}+{\bf c}_i\Delta t, t + \Delta t) = \bar f_i({\bf x},t) +
R_i({\bf x}, t) \Delta t.
\end{equation}

This shows that the LBE evolution can be thought of two separate
processes: the first is a relaxational step in which the distributions
$\bar f_i$ are relaxed to their ``post-collisional'' values $\bar
f_i({\bf x},t^{\ast})$,
\begin{equation}
\bar f_i({\bf x},t^{\ast})=\bar f_i({\bf x},t)
+ R_i({\bf x}, t) \Delta t,
\end{equation}
followed by a propagation step in which the post-collisional
distributions are propagated along a Lagrangian trajectory without
further change,
\begin{equation}
\bar f_i({\bf x}+{\bf c}_i\Delta t, t + \Delta t) = \bar f_i({\bf x},t^{\ast}).
\end{equation}
Thus the computational part of the method is most naturally framed in
terms of the auxiliary distributions $\bar f_i$ and not the physical
distribution functions $f_i$ themselves. To obtain the
post-collisional $\bar f_i$ without having to refer to the $f_i$, the
latter must be eliminated from Eq.~(\ref{halftransform}). Inverting
the equations defining the $\bar{f}_i$ in Eq.~(\ref{transformf}) we
obtain
\begin{equation}
R_i = ({\mathbf 1}+\frac{\Delta t}{2}{\cal L})^{-1}_{ij}
\left[- {\cal L}_{jk} ( \bar f_k-f_k^0) + \Phi_j({\bf x},t)\right].
\end{equation}
Combining this with Eq.~(\ref{halftransform}) we obtain a numerical
scheme for the forced discrete Boltzmann equation with a general
collision operator in terms of the $\bar f_i$:
\begin{widetext}
\begin{equation}
\label{mrtlbe}
\bar f_i({\bf x}+{\bf c}_i\Delta t,t+\Delta t)=
\bar f_i({\bf x},t) + ({\mathbf 1}+\frac{\Delta t}{2}{\cal L})^{-1}_{ij}
[- {\cal L}_{jk} (\bar f_k - f_k^0) + \Phi_j({\bf x},t)].
\end{equation}
For a single relaxation time collision operator, where ${\cal
L}_{ij}=\delta_{ij}/\tau$, this takes on a particularly simple form
\begin{equation}
\label{bgklbe}
\bar f_i({\bf x}+{\bf c}_i\Delta t,t+\Delta t) = 
\bar f_i({\bf x},t) 
+ \frac{\Delta t}{\tau + \frac{\Delta t}{2}}\left[ -(\bar f_i - f_i^0) + 
\tau \Phi_i({\bf x},t) \right],
\end{equation}
\end{widetext}
a result obtained previously by a multiscale expansion of the LBE
dynamics \cite{Guo:2002}. For a non-diagonal collision operator, the
collision term is best evaluated in the moment basis. For example,
using a collision operator in which the ghost modes are projected out
and the stress modes relax at a rate $\tau^{-1}$, the post-collisional
$\bar f_i$ ({\em i.e.} the RHS of Eq.~(\ref{mrtlbe})) is given by
\begin{equation}\label{fiexp}
  \bar{f}_i({\bf x},t^{\ast}) = w_i\left[
    \rho + 
    \frac{A_\alpha c_{i\alpha}}{c_s^2} +
    \frac{B_{\alpha\beta} Q_{i\alpha\beta} }{2c_s^4}
  \right],
\end{equation}

where $A_\alpha$, the momentum component of the post-collisional
auxiliary distributions, is
\begin{equation}
A_\alpha = \sum_{i=0}^n \bar{f}_i c_{i\alpha} + \rho F_\alpha \Delta t 
\end{equation}
and $B_{\alpha\beta}$, the stress component, is
\begin{eqnarray}
&& B_{\alpha\beta} = \sum_{i=0}^{n} \bar f_i Q_{i\alpha\beta} +
  \nonumber \\ &&\frac{\Delta t}{\tau + \Delta t /2}
  \left(\sum_{i=0}^{n} \bar f_i Q_{i\alpha\beta} - \rho v_\alpha
  v_\beta + \tau (v_\alpha F_\beta + F_\alpha v_\beta) \right)
\end{eqnarray}

The hydrodynamic variables are moments of the physical distribution
$f_i$, but can easily be obtained from the auxiliary distributions
$\bar f_i$ used in the computation, using the transformation rule,
Eq.~(\ref{transformf}), the definitions of the macroscopic variables,
Eq.~(\ref{macrovariable1}), and the constraints of mass and momentum
conservation, Eq.~(\ref{conservation}). We obtain
\begin{widetext}
\begin{subequations}
\begin{eqnarray}\label{macrovariables2}
  \rho &=& \sum_{i=0}^{n}\bar f_i,\quad \\ \rho v_\alpha &=&
  \sum_{i=0}^{n}\bar{f}_i c_{i \alpha} + \rho F_\alpha\frac{\Delta
  t}{2},\quad \\ S_{\alpha\beta} &=& \sum_{i=0}^{n} \bar f_i
  Q_{i\alpha\beta} + \frac{\Delta t/2}{\tau + \Delta t /2}
  \left(\sum_{i=0}^{n} \bar f_i Q_{i\alpha\beta} - \rho v_\alpha
  v_\beta + \tau (v_\alpha F_\beta + F_\alpha v_\beta) \right)
\end{eqnarray}
\end{subequations}
\end{widetext}

The equilibria can be reconstructed from $\rho$ and $\rho
\vec{v}$. What appear in the literature as redefinitions of momentum
and stresses are shown in the above analysis to be discretization
errors which vanish as $\Delta \rightarrow t 0$. This completes the
description of the method for the numerical solution of the forced
LBE.  Verberg and Ladd have derived results {\altered equivalent} to
those above using a multiple scale analysis of the discrete LBE
dynamics \cite{Ladd:2001}. {\altered It is not clear to us whether
their analysis admits singular force densities. However, the above
derivation shows that these equations are a reliable starting point
even in that case.}

The LBE can be extended to situations where the fluctuations in the
fluid density and momentum are important \cite{Ladd:1994}. A
consistent discrete kinetic theory of fluctuations was presented in
\cite{Adhikari:2005}, which improves on an earlier algorithm due to
Ladd \cite{Ladd:1994}, and produces thermodynamically {\altered
accurate} variances of the local mass and momentum
densities. {\altered We return to the issue of noise below, when we
address the representation of Brownian colloids as point particles
(Section \ref{sec:fallers}).}

\section{Singular force densities}\label{sec:singular}
In a wide variety of situations, as mentioned in the Introduction,
force densities may need to be defined off-lattice, and may in
addition be singular. Mathematically, such a force density may be
written as
\begin{equation}\label{singularF}
{\bf F}({\bf r}) = \int {\bf f (R)}\delta({\bf r -R})d\lambda
\end{equation}
where the force is localized to some manifold described parametrically
as ${\bf r = R}(\lambda)$ and $d\lambda$ is the measure on the
manifold. Any numerical method which attempts to deal with such force
distributions must be reconciled with the singular nature of the force
and, for grid-based numerical methods, the fact that the position of
the manifold need not coincide with the nodes of the grid. In a well
established numerical method \cite{Peskin:2002}, the Dirac delta
function in the singular force distribution is replaced by a
regularized delta function which leads to a smooth distribution of
forces. Necessarily, this implies that the force is now no longer
localized on the manifold but is sharply peaked and smooth around
it. This smooth force density can now be sampled on the grid using the
discretized delta function as an interpolant. Thus a representation of
Eq.~(\ref{singularF}) on the grid is obtained from
\begin{equation}
{\bf F}({ \bf r}) = \sum_{a}{\bf f}({\bf R}_{a})\delta^{P}({\bf r - R}_a)
\end{equation}
The crucial ingredient here is the kernel function $\delta^P$, which
is a representation of the Dirac delta function regularized on the
grid.  We have followed closely the method described by Peskin
\cite{Peskin:2002} where a regularized approximation to the Dirac
delta function with compact support is derived:
\begin{equation}
\delta^P({\bf r}) = \frac{1}{h^3} \, f\left(\frac{x}{h}\right)
f\left(\frac{y}{h}\right)
f\left(\frac{z}{h}\right)
\end{equation}
where $h=\Delta x =\Delta y = \Delta z$ is the lattice spacing and
$f(r)$ is given by:
\begin{equation}\label{deltadef}
f(r) = \begin{cases}
  \frac{3 - 2|r| + \sqrt{1 + 4|r| - 4r^2} }{8} & |r| \leq 1, \cr
  \frac{5 - 2|r| - \sqrt{-7 + 12|r| - 4r^2}}{8} & 1 \leq |r| \leq 2, \cr
  0 & |r| \geq 2.\cr
\end{cases}
\end{equation}
This form is motivated by the need to preserve the fundamental
properties of the Dirac delta function on the grid
\cite{Peskin:2002}. A simple closed form approximation to $\delta^P$
which is useful for analytical work is
\begin{equation} \label{deltaapprox}
  \tilde{ f }(r) = \begin{cases} \frac{1}{4}\left(1 + \cos
    \left(\frac{\pi r}{2} \right)\right) & |r| \leq 2,\cr 0 & |r| > 2,
  \end{cases}
\end{equation}
whose Fourier transform is given by:
\newcommand{\sinc}{\hbox{sinc}}
\begin{equation}\label{deltaFT}
  \tilde{ f }(k) = \sinc 4\pi k + \frac{1}{2} \sinc \left(4 \pi k -
  \pi\right) + \frac{1}{2} \sinc \left(4 \pi k + \pi\right).
\end{equation}
{\altered In this work we combine Eq.~(\ref{singularF}) directly with
the numerical method described in the previous section, giving a
well-defined method for incorporating singular and/or off-lattice
force densities into the lattice Boltzmann hydrodynamics.}

\subsection{Validation}
\begin{figure}
\centering
\includegraphics[width=0.5\textwidth]{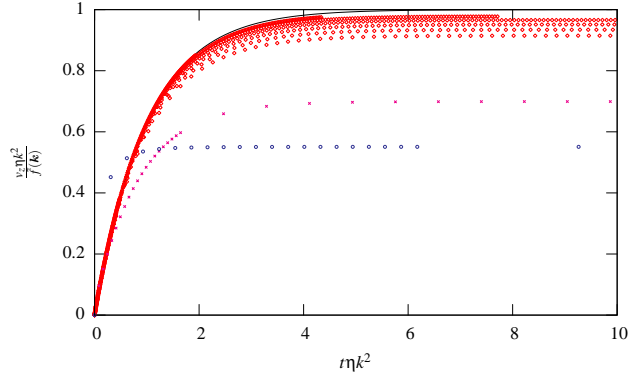}

\caption{Relaxation of the solenoidal component of the velocity due to
a delta function forcing. The simulations were performed on a $64^3$
lattice. Shown are the first $8$ (upper curves), the $16$th (middle
curve) and the $31$st (lower curve) Fourier modes of the velocity
field. The solid line is the analytical result,
Eq.~(\ref{eqn:oseen}).}
\label{fig:oseen}
\end{figure}

To validate the method, we compare analytical solutions of the
singularly forced Navier-Stokes equation against our numerical
solutions, using lattice units $\Delta x = \Delta t = 1$, $\rho = 1$.
The most straightforward benchmark is against the initial value
problem for the Stokes limit
\begin{equation}
\partial_t \vec{v} = -\nabla p + \eta\nabla^2\vec{v} +   \vec{F}(\vec{r})
\end{equation} 
where the nonlinearity has been discarded, incompressibility is
assumed, and $\vec{F}(\vec{r}) = \vec{F}_0 \delta
(\vec{r}-\vec{R}_0)$. In an infinite system, the solution is obtained
in terms of the unsteady Oseen tensor describing the diffusion of
vorticity \cite{Landau:1971}. In a system with periodic boundary
conditions, the Oseen solution must be replaced by the Hasimoto
solution \cite{Hasimoto:1959}. In contrast to the Oseen solution, the
real-space Hasimoto solution is not available in a simple closed form
but must be evaluated numerically. However, the solution in Fourier
variables presents no such difficulty, and is in fact identical in
both cases:

\begin{equation}\label{eqn:oseen}
  {\bf {v}}({\bf k},t) = \frac{1 - e^{-\eta k^2 t}}{\eta k^2}(\eye -
  {\bf \hat k \hat k})\cdot{\bf F}({\bf k})
\end{equation}
Thus, we find it most convenient to compare Fourier modes of the
velocity from the numerical solution against the solution above. In
particular, this provides a neat way to evaluate the performance of
the method at different length scales.  In FIG.~\ref{fig:oseen} we
compare the numerical data (points) to the theoretical result for a
regularized force monopole using the approximation to the Peskin delta
function, Eq.~(\ref{deltaFT}) (solid line).

The results show excellent agreement with the theoretical curve for
low $k$ modes, where we expect the momentum to behave
hydrodynamically. The departure from hydrodynamic behaviour increases
progressively with the wavenumber, as expected from previous studies
on the hydrodynamic behaviour of the LBE
\cite{Lallemand:2000}. However, there is a significant range of length
scales over which our model reproduces hydrodynamic behaviour, which
is not less than the scale over which hydrodynamic behaviour is
obtained in the unforced LBE~\cite{PhysRevE.50.4586}.

\begin{figure}
  \subfigure[]{\label{fig:stresslet-comp}
    \centering
    \includegraphics[width=0.5\textwidth]{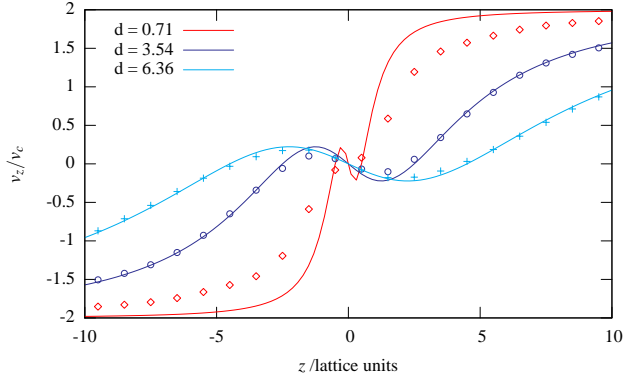}
  }
  \subfigure[]{\label{fig:stresslet-diff}
    \centering
    \includegraphics[width=0.5\textwidth]{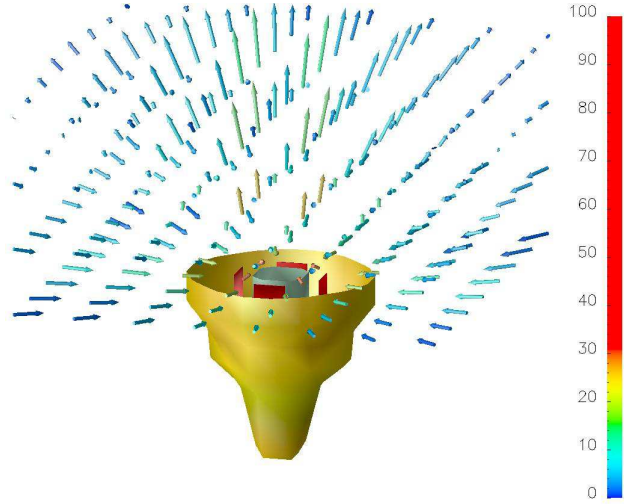}
  }
  \caption{(Colour online) Velocity around a symmetric point-force
  dipole, normalized by a characteristic speed for that distance from
  the dipole, $v_c \equiv \frac{Fa}{8\pi\eta
  r^2}$. FIG.~\ref{fig:stresslet-comp} shows velocity along lines
  parallel to the forces, at several separations. Points are
  simulation results, lines theoretical predictions with no free
  parameters. In figure~\ref{fig:stresslet-diff}, the upper half shows
  the simulated velocity field. The lower half shows isosurfaces of
  the magnitude of the velocity difference between simulation and
  theory at values of 25\% and 50\%. The colouring (online) depends
  upon the magnitude of the difference field and is shown as a
  percentage of $v_c$ in the colour bar. The force dipole is oriented
  vertically and positioned in the centre of the volume.}
\label{fig:stresslet}
\end{figure}

\begin{figure}
  \centering
  \includegraphics[width=0.5\textwidth]{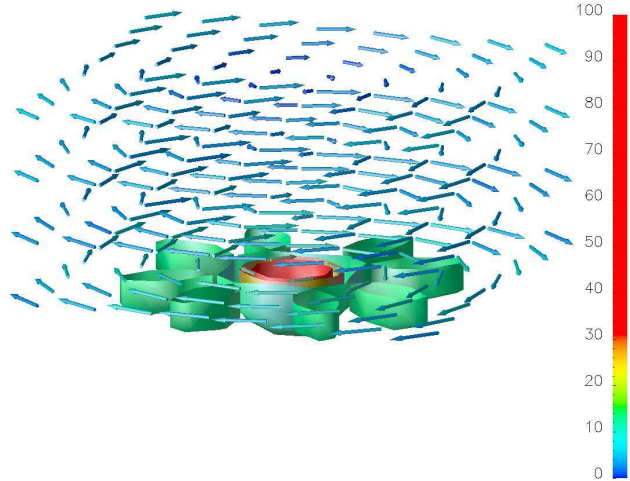}
  \caption{(Colour online) Velocity field around a regularized rotlet,
  normalized by $v_c$ (see FIG.~\ref{fig:stresslet}). Upper half:
  simulated velocity field. Lower half: isosurfaces of the magnitude
  of the velocity difference between simulation and
  theory. Isosurfaces are at values of 12.5\%, 25\% and 37.5\%.  The
  colouring (online) depends upon the magnitude of the difference
  field and is shown as a percentage of $v_c$ in the colour bar. The
  rotlet is oriented with the forces in a horizontal plane and
  positioned in the centre of the volume.}
  \label{fig:rotlet}
\end{figure}

By combining elementary monopoles, discrete representations of higher
multipoles can be generated.  For example, the discrete Stokes
doublet, a dipole of two point forces, can be constructed out of
monopoles of magnitude $F$ and separation $a$ and is often used as a
simplified representation of a neutrally buoyant, steadily moving
self-propelled particle.  In figures \ref{fig:stresslet-comp} and
\ref{fig:stresslet-diff}, we compare the velocity response {\altered
of such a dipole} to theoretical predictions, finding good agreement
away from the immediate vicinity of the forces.  In
FIG.~\ref{fig:rotlet}, we show a velocity field plot for the
antisymmetric force dipole, or rotlet, which may be used as a
representation of an object which rotates due to an external torque.
{\altered This requires the use of four, rather than two, point
forces; we arrange these in a swastika-like fashion (whose axes can be
aligned in an arbitrary direction without significantly affecting the
flow produced). This cancels a spurious stresslet component that
arises from our regularization of the $\delta$ function for any dipole
in which the forces are not collinear with the separation vector.}
 
{\altered The above examples show that the regularized delta function
provides an useful way of incorporating arbitrary distributions of
singular forces into the lattice Boltzmann method, capable of dealing
with internal as well as external forcing.}

\section{A Stokeslet model for dilute colloids}\label{sec:fallers}
The dynamics in a dilute sedimenting suspension, despite a century of
investigation, still presents open questions \cite{Ladd:2001}. The
problem, even for a hard-sphere suspension, is unusually difficult due
to the long-ranged, many-body nature of the hydrodynamic
interaction. Moreover, the flow can develop structural features at
large length scales, and the role of inertia, while usually negligible
at the particle scale, may be significant at those scales
\cite{Segre:1997}. The Stokes approximation of globally vanishing
Reynolds number cannot thus be justified \emph{a priori} in a
sedimenting suspension. The full hydrodynamic problem including
inertia for both fluid and particles was first simulated by Ladd using
a novel lattice Boltzmann method \cite{Ladd:1993}. This method, though
possibly the most competitive for fully resolved particles, remains
computationally expensive. A considerable simplification of the
hydrodynamics is possible if only the lowest order multipole of the
force distribution induced on the particle surface by the no-slip
boundary condition is retained. This principle was exploited
previously to develop representations of polymers as strings of point
particles which were then coupled to an LB fluid \cite{usta:2005,
Ahlrichs:1998}. A similar idea has been used \cite{lobaskin:2004a,
lobaskin:2004b} to represent resolved colloids with a mesh of point
particles covering their surfaces. However in the current work we
simplify further, treating each colloid as a single point particle
(thereby sacrificing all near-field effects).In the colloidal context,
this model was first introduced by Saffman \cite{Saffman:1973}; the
finite sized particles are replaced by a singular force monopole, the
Stokeslet, located at the nominal centre of the particle. In Saffman's
original model, both the fluid and the particles have no inertia. In
keeping with the comments above, our model retains inertia for the
fluid, while neglecting it for the particle and for hydrodynamics at
the particle scale. We thus have a momentum balance equation,
\begin{equation}
  \rho\partial_t{\bf v} = -\nabla p + \eta\nabla^2{\bf v} + \sum_{s}
	      {\bf F}_{s}\delta({\bf r - R}_{s})
\end{equation}
where the sum includes contributions from the $s = 1\ldots N$
particles located ${\bf R}_s$ and acted upon by \emph{external} forces
${\bf F}_s$. In the absence of particle inertia accelerations vanish,
and the particle coordinates are updated directly using the first
Fax\'en relation \cite{pozrikidis92}
\begin{equation}
  \dot{\bf R}_{s} = {\bf v}^{\infty}({\bf R}_{s}) + \frac{{\bf
      F}_{s}}{6\pi\eta a},
\end{equation}
which relates the centre of mass velocity of the particle $\dot{\bf
R}_s$ to the external force on it ${\bf F}_s$.  The background
velocity ${\bf v}^{\infty}({\bf R}_s)$ is the fluid velocity at the
location ${\bf R}_s$ \emph{in the absence} of the $s$-th particle. The
above two equations provide a complete specification of a model of
sedimenting spheres, valid in the dilute limit, for dynamics at long
wavelengths.

The lattice Boltzmann implementation of this model proceeds by first
replacing the Dirac delta function with the regularized delta
functions to obtain a force density at the grid points ${\bf F}({ \bf
r}) = \sum_{s}{\bf F}_s\delta^{P}({\bf r - R}_s)$.  Since the LBE
evolves the total fluid velocity ${\bf v}({\bf r})$ due to all
particles, the background fluid velocity ${\bf v}^{\infty}$ must be
obtained by a careful subtraction procedure. In the absence of fluid
inertia at the particle scale this can be accomplished as follows. By
definition, the fluid velocity at a node ${\bf v}({\bf r})$ is the sum
of the background velocity at the node ${\bf v}^{\infty}({\bf r})$ and
the velocity due to the $s$-th Stokeslet located at ${\bf R}_s$, ${\bf
v}({\bf r}) = {\bf v}^{\infty}({\bf r}) + {\bf v}_{s}({\bf r}, {\bf
R}_s)$. The background velocity field at the location of the particle
can be obtained using the same interpolation kernel as used for the
force, ${\bf v}^{\infty}({\bf R}_s) = \sum_{\bf r}{\bf
v}^{\infty}({\bf r})\delta^{P}({\bf R_s - r})$, and using the previous
relation can be written as
\begin{equation}
  {\bf v}^{\infty}({\bf R}_s) = {\bf v}({\bf R}_s) - \sum_{\bf r}{\bf
    v}_{s}({\bf r},{\bf R}_s)\delta^{P}({\bf R_s - r})
\end{equation}
Appealing only to linearity and dimensional analysis, the sum above
can be expressed as
\begin{equation}\label{eq:a_Ldefn}
  \sum_{\bf r}{\bf v}_{s}({\bf r},{\bf R}_s)\delta^{P}({\bf R_s -
    r})\equiv \frac{{\bf F}_s}{6\pi\eta a_L({\bf R}_s)}
\end{equation}
In Appendix~\ref{sec:subtraction}, we derive this result and show that
the lattice parameter $a_L$ depends only on the system size $L$ and on
the form of the regularization and interpolation kernels; it is
independent of viscosity $\eta$ and of the radius $a$.  Using
Eq.~(\ref{eq:a_Ldefn}), the update equation for the Stokeslet
positions can now be expressed in terms of the interpolated fluid
velocity, without any reference to the background velocity,
\begin{equation}
  \dot{\bf R}_{s} = {\bf v}({\bf R}_{s}) + \frac{{\bf
      F}_{s}}{6\pi\eta}\left(\frac{1}{a}-\frac{1}{a_L({\bf R}_s)}\right)
\end{equation}
Notice that replacing the background velocity in the Fax\'en relation
with the actual fluid velocity induces an effective backflow, leading
to a renormalized hydrodynamic radius,
\begin{equation}
\frac{1}{a_{R}} = \frac{1}{a} - \frac{1}{a_L}
\end{equation}
The numerics thus places a constraint $a \ll a_L$ on the allowed
values of the hydrodynamic radius $a$. This numerical constraint
encodes the condition that the grid points must be in the far-field of
the Stokeslet, the limit in which the background velocity can be
obtained from the fluid velocity by subtracting a monopole
contribution. In our simulations, we operate well within this limit.

This almost completes the description of the lattice Boltzmann
implementation of our Stokeslet model of sedimenting particles. The
only free parameter is the hydrodynamic radius $a$ of the particles,
which decides how fast they sediment for a given force ${\bf F}_s$. As
shown below, the lattice parameter $a_L$ can be calculated
analytically as a function of system size. We find it convenient to
fit it using a procedure described in
Appendix~\ref{sec:subtraction}. {\altered Finally, to address Brownian
motion of our colloids, we need to use the FLBE of
\cite{Adhikari:2005} which imparts an appropriate thermal noise
spectrum to the fluid. Because of the renormalization of $a$, the
resulting diffusivity is generally not correct unless a further noise
term is added that is the counterpart of the $a_L$ correction. The
details are explained in Appendix~\ref{sec:noise}.}

\subsection{Benchmarks}
\begin{figure}
  \centering
  \includegraphics[width=0.5\textwidth]{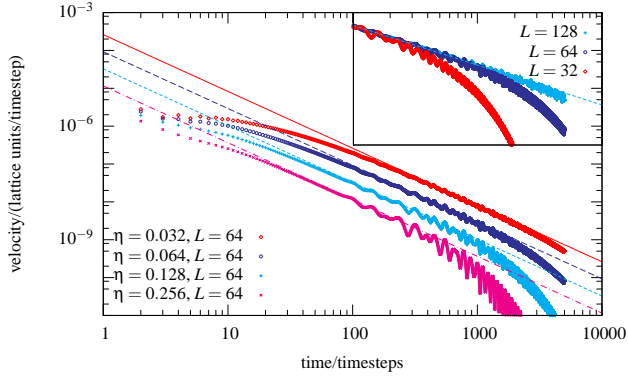}
    
  \caption{(Colour online) Response of a single particle to an
impulsive force $\vec{P}$ in a periodic box. Main figure: response
shown for a range of viscosities (points) compared to the theoretical
prediction $\vec{v} = \frac{\vec{P}}{12\rho (\pi \nu t)^{-3/2}}$ at
long times \cite{Landau:1971}. The inset shows the effect of varying
the size of the simulation box. Deviations from the prediction become
significant at approximately 250, 1000 and 4000 timesteps for box
sizes of 32, 64 and 128, respectively. This is consistent with the
expected scaling, $\tau \sim L^2/\eta$.}
\label{fig:tail}
\end{figure}

Our first benchmark addresses the dynamics of a single impulsively
started particle, without noise. From unsteady hydrodynamics, we know
that the asymptotic decay of the particle velocity varies as
$t^{-d/2}$ in $d$ dimensions \cite{Landau:1971}. In
FIG.~\ref{fig:tail} we display the decay of the particle velocity, for
a single hydrodynamic radius (0.05~LU), but several values of the
fluid viscosity.  In all cases, we see the correct asymptotic
behaviour, until the particle is begins to interact with its image,
due to the periodic boundary conditions.  {\altered (This
interpretation is supported by the scaling of the time $\tau$ at which
the deviations become significant: we find $\tau \simeq
\frac{L^2}{\eta}$ as shown in the inset to FIG.~\ref{fig:tail}.)}  In
other words, our particle model correctly captures the low and
intermediate frequency behaviour of the particle mobility, but cannot
capture the high-frequency behaviour correctly, since that depends on
the way vorticity diffuses in the immediate neighbourhood of the
particle, a regime which is excluded in our model.

\begin{figure}
  \centering
  \includegraphics[width=0.5\textwidth]{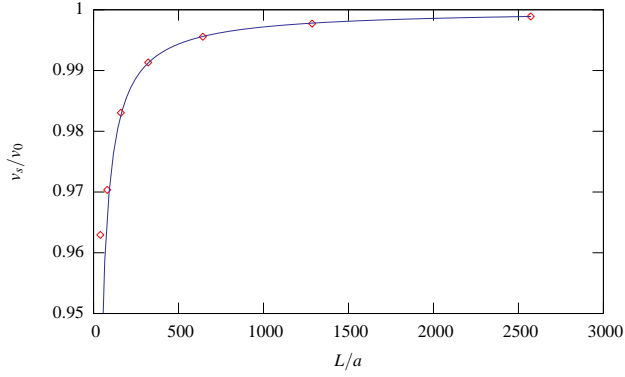}%
  
  \caption{(Colour online) Steady sedimentation velocity of a simple
cubic array of spheres, normalized by the Stokes sedimentation
velocity of a single particle, $v_0$. The separation, $L$, is
expressed in terms of the fitted particle radius, $a$. The solid line
is the theoretical result $v = v_0/(1 + k\frac{a}{L})$, with $k =
2.84$ \cite{happelbrenner,Hasimoto:1959}.}
  \label{fig:lattice}
\end{figure}

\begin{figure}
\centering
\subfigure[$\; t = 0 t_\mathrm{St}$]{
  \includegraphics[width=0.3\textwidth]{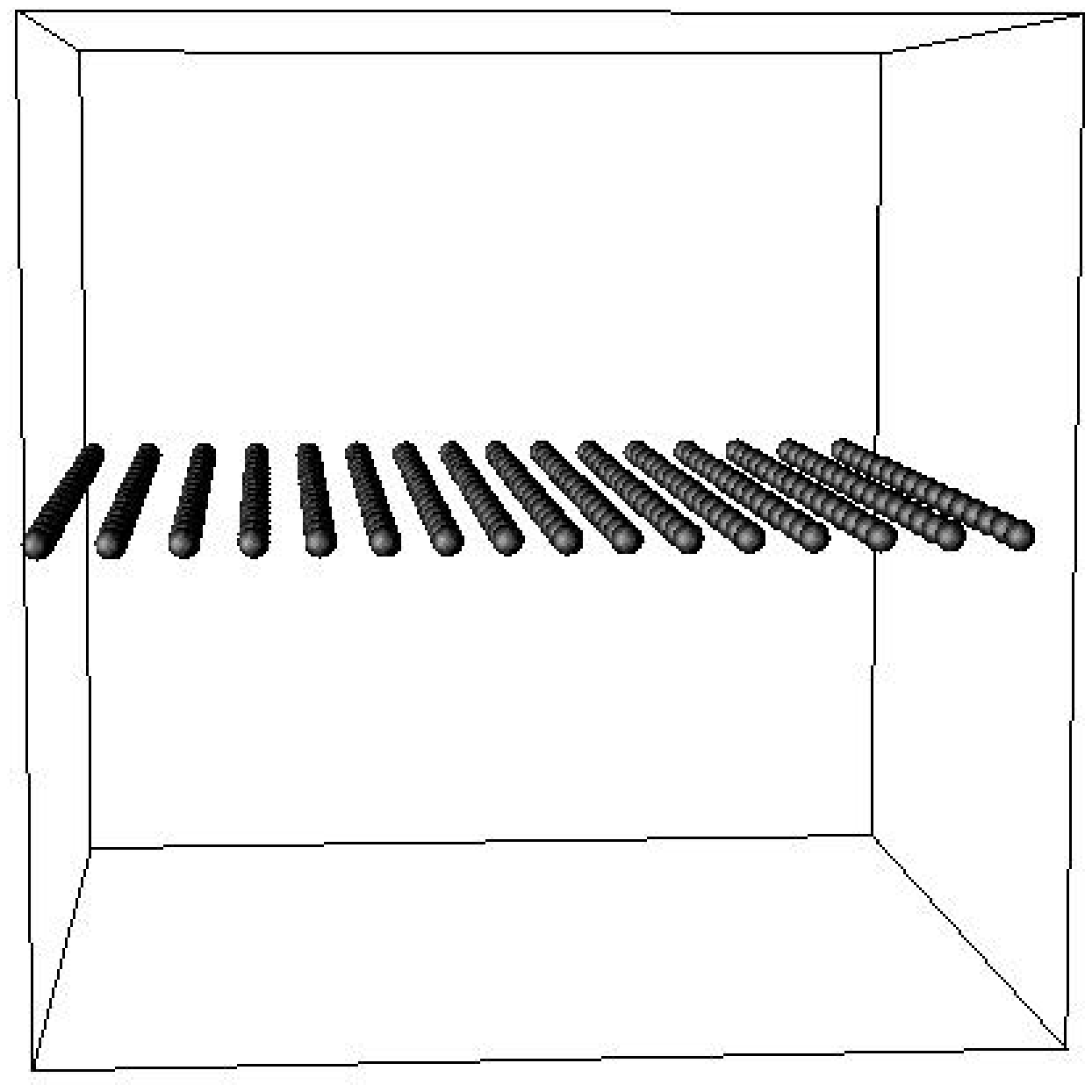}
}
\subfigure[$\; t = 1650 t_\mathrm{St}$]{
  \includegraphics[width=0.3\textwidth]{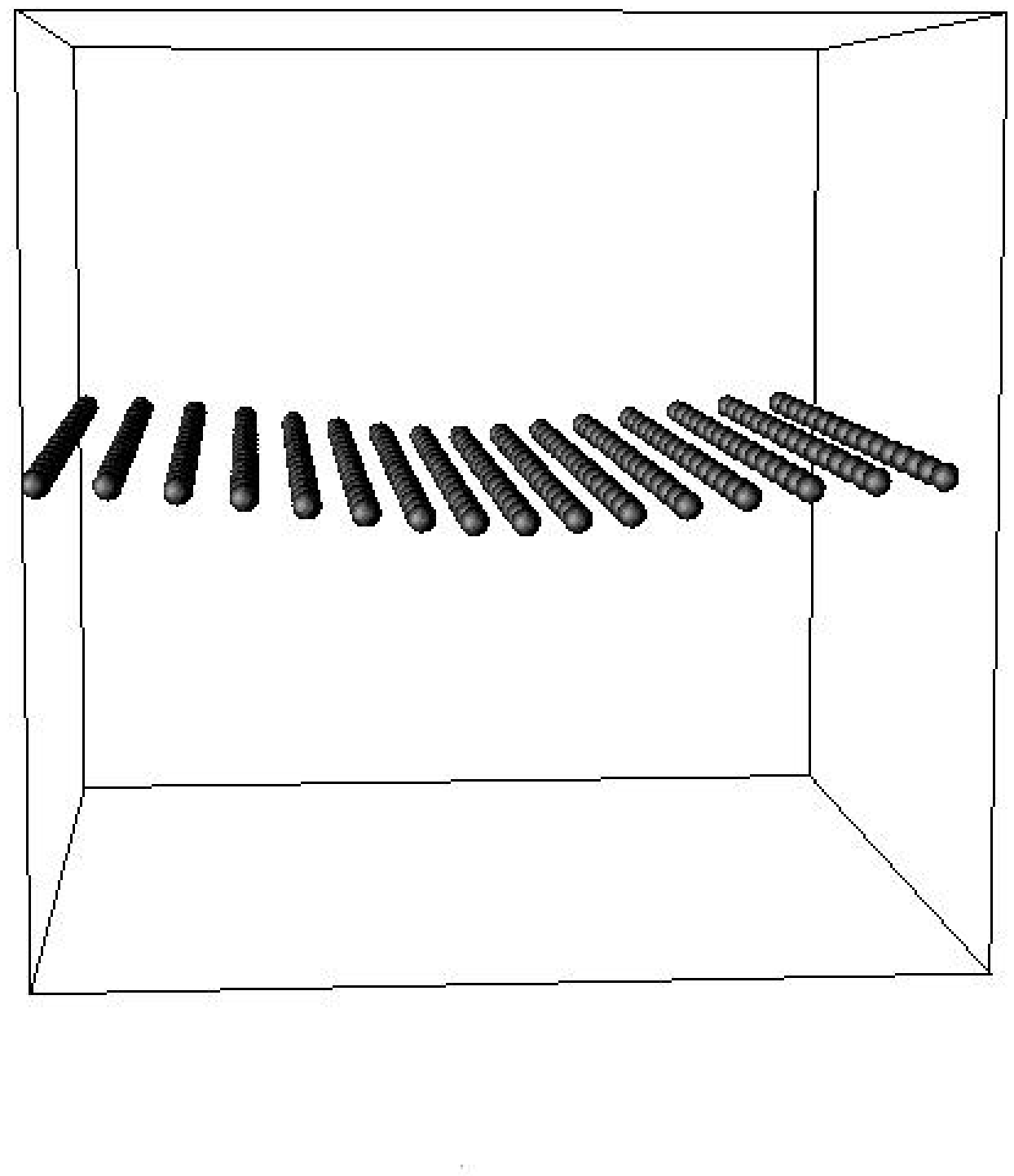}
}
\subfigure[$\; t = 3300 t_\mathrm{St}$]{
  \includegraphics[width=0.3\textwidth]{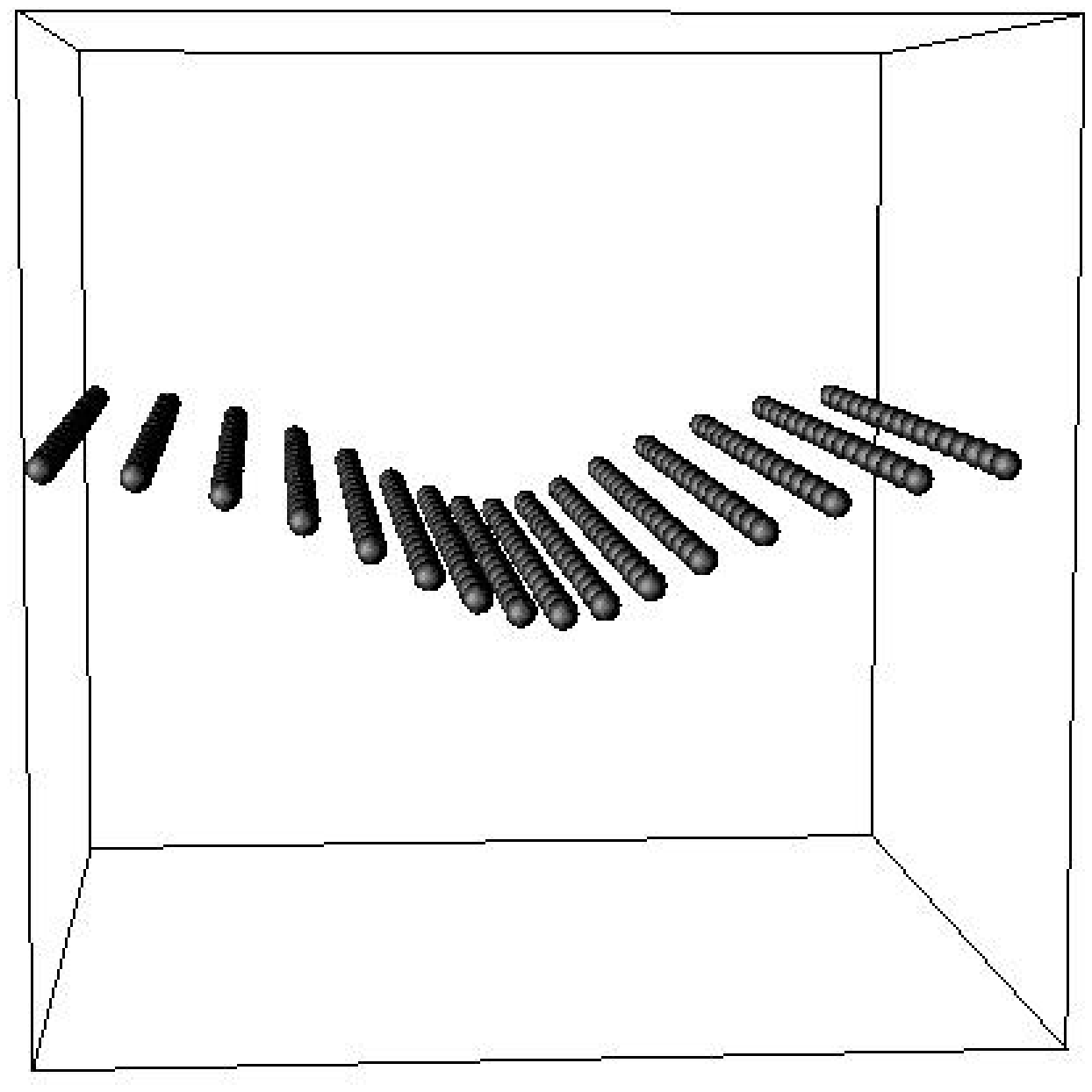}
}

\caption{Crowley instability of a 2-dimensional lattice of sedimenting
particles. Images are generated in the comoving frame and particle
size has been exaggerated for illustrative purposes.}
\label{fig:crowley}
\end{figure}

Our next benchmark involves collective motion of a set of particles,
and thus directly probes the hydrodynamic interaction between
particles. In FIG.~\ref{fig:lattice} the mean sedimentation velocity
of a periodic array of spheres is shown, as a function of volume
fraction. There is excellent agreement with the theoretical result of
\cite{happelbrenner}.
The model is also able faithfully to capture instabilities due to
collective hydrodynamic flow. In FIG.~\ref{fig:crowley} the
instability of a falling 2D lattice of spheres in three dimensions
\cite{crowley:1296} is captured, at least qualitatively, by our model.

\begin{figure}
\subfigure[ $\; \mathrm{Re}=10^{-6}$]{\label{fig:reynoldsRangeRef}
  \includegraphics[width=0.3\textwidth]{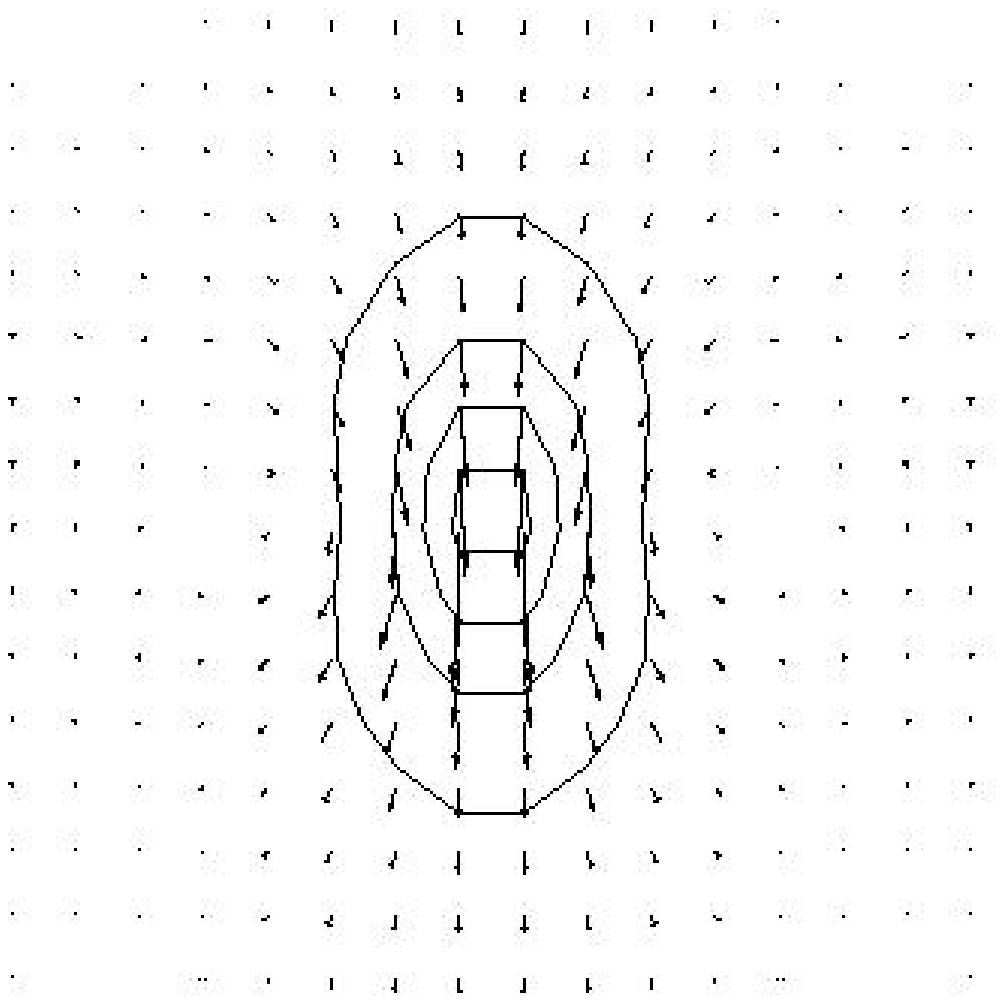}
}
\subfigure[ $\; \mathrm{Re}=10^{-4}$]{\label{fig:reynoldsRangeDiff4}
  \includegraphics[width=0.3\textwidth]{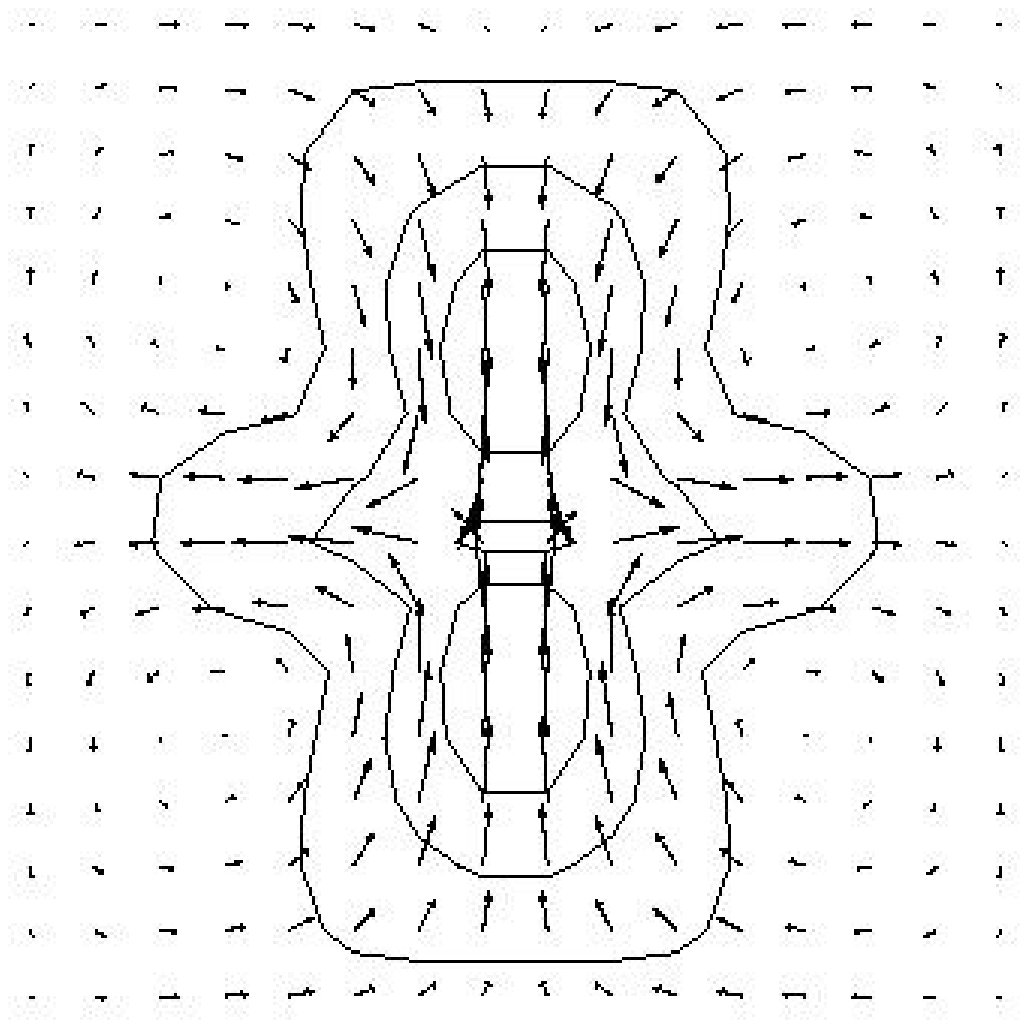}
}
\subfigure[ $\; \mathrm{Re}=10^{-2}$]{\label{fig:reynoldsRangeDiff2}
  \includegraphics[width=0.3\textwidth]{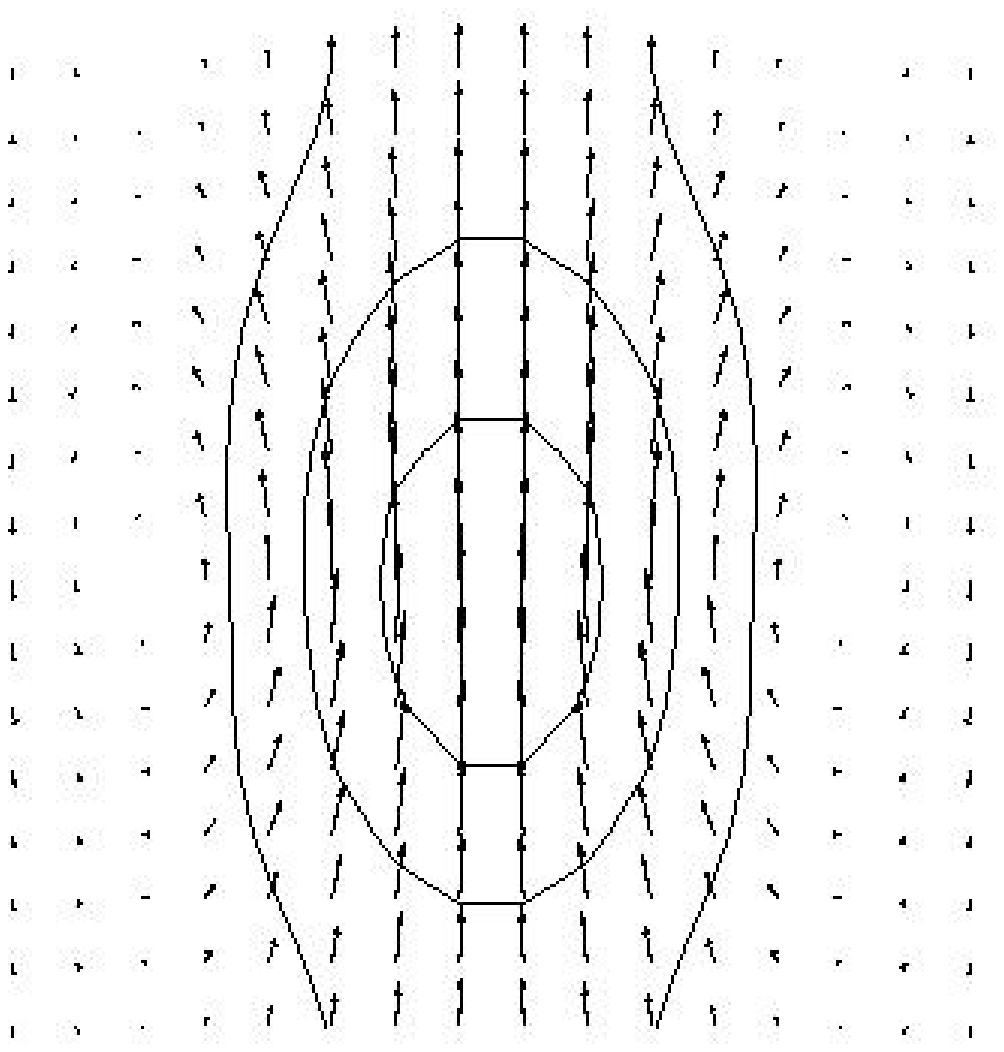}
}
\caption{Contour plots of the normalized velocity field $\vec{u}$ for
a reference simulation at very low $\mathrm{Re} = 10^{-6}$ (top); and
velocity difference fields for $\mathrm{Re} = 10^{-4}$ (middle) and
$10^{-2}$ (bottom). These are for a point-like colloid sedimenting in
a $32^3$ box with periodic boundary conditions. Reference case:
contour interval $0.02 v_0$. $\mathrm{Re} = 10^{-4}$ case: contour
interval $5\times 10^{-6}$. $\mathrm{Re} = 10^{-2}$ case: contour
interval $5\times 10^{-3}$.}\label{fig:reynoldsRange}
\end{figure}

For most problems, all Rey\-nolds numbers below some
(sit\-uation-dependent but small) value give rise to equivalent
behaviour, as discussed in detail in previous work~\cite{Cates:2004}.
Following protocols discussed there, we have compared the normalized
velocity field $ \vec{u} = \vec{v} / v_0$ (with $v_0$ the
sedimentation velocity of an isolated colloid) for a number of
simulations of a single sedimenting sphere with periodic boundary
conditions (see figure~\ref{fig:reynoldsRange}) in order to explore
the range of Reynolds number at which our algorithm gives acceptably
accurate results.  Our `reference' simulation has a very small
$\mathrm{Re} = 10^{-6}$ such that we can be confident it is the in the
Stokesian limit~\cite{Cates:2004}.  This is shown in the panel
\ref{fig:reynoldsRangeRef}.  Panels \ref{fig:reynoldsRangeDiff4} and
\ref{fig:reynoldsRangeDiff2} show the normalized velocity difference
fields between the reference case and simulations with $\mathrm{Re} =
10^{-4}$ and $\mathrm{Re} = 10^{-2}$, respectively.  In the simulation
with $\mathrm{Re} = 10^{-4}$, the magnitude of the difference is
everywhere less than $2\times10^{-5}$, a negligibly small error.  In
the simulation with $\mathrm{Re} = 10^{-2}$, we find $|\Delta \vec{u}|
\le 0.01$ throughout the bulk of the domain; only in the immediate
vicinity of the particle does it become larger.  This suggests that
this Reynolds number is sufficiently low to give `realistic', although
not `fully realistic', behaviour \cite{Cates:2004}.  Since reaching
very low Reynolds number requires paying a larger cost in
computational time, and there are other sources of percent-level error
in the code, $\mathrm{Re}=10^{-2}$ is probably a reasonable compromise
between accuracy and run-time, for studies in the low Reynolds number
limit.

\subsection{Comparision to a fully resolved LB algorithm}
\begin{figure}
  \includegraphics[width=0.8\textwidth]{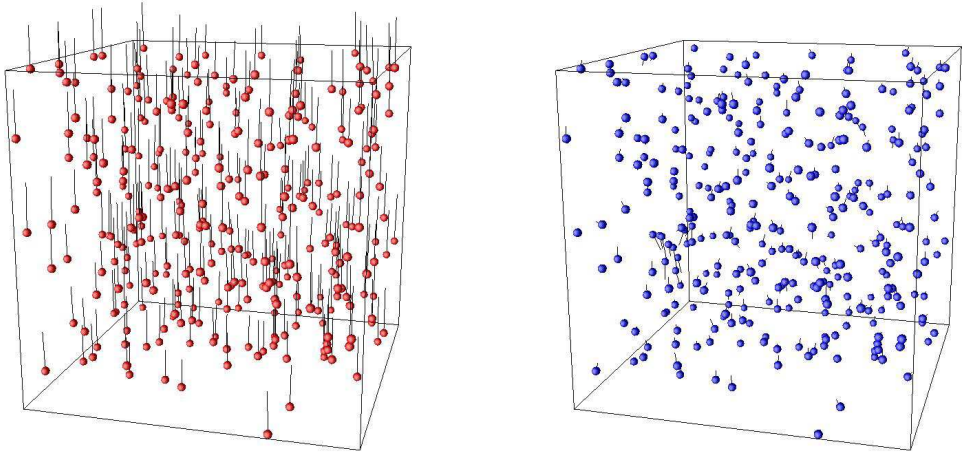}
  \caption{(Colour online) Both visualizations show the particles at
    their position in the point-particle simulation after 10 Stokes
    times.  Left: lines show the trajectories from the starting
    configuration.  Right: lines show the difference between
    fully-resolved and point like algorithm.  Parameters for the two
    systems are shown in Table~\ref{tab:params}. See also Movie 1 \cite{supmat}.}
  \label{fig:random-traj-diff}
\end{figure}

\begin{figure}
  \centering
  \includegraphics[width=0.5\textwidth]{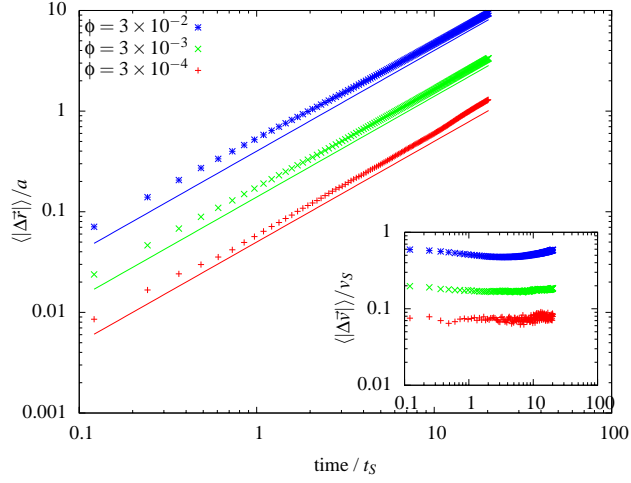}
  \caption{(Colour online) Main figure: mean absolute difference in
    position between fully-resolved and point-like sedimenting
    particles. Inset: mean absolute difference in velocity between the
    two simulations. Parameters for the two systems are shown in
    Table~\ref{tab:params}.}
  \label{fig:deltaR}
\end{figure}

As a final benchmark, we have compared the behaviour of our
sedimenting particle model with a fully resolved colloid simulation
code using the algorithm of Nguyen and Ladd~\cite{PhysRevE.66.046708}.
{\alteredRWN (For full implementation details see
\cite{Stratford:2005a}.)}  {\altered At dilute concentrations, the
paths of the particles are almost indistinguishable between the two
simulations when plotted graphically}. This is shown in
figure~\ref{fig:random-traj-diff} {\alteredRWN for volume fraction
$\phi = 3.0 \times 10^{-3}$}; note that the largest differences occur
when the density of particles is large locally, when the implicit
assumption of our model that the particles are always at separations
large compared to their radius is no longer valid.  We cannot expect
both simulations to give the same trajectories for long times, since
the small differences between algorithms will cause exponential
separation of trajectories owing to the positive Lyapunov exponent of
the system. However, from a plot of the mean difference in position
between the two simulations against time, we can see {\altered
excellent agreement for several Stokes times, and until at least ten
Stokes times for sufficiently dilute systems (FIG.~\ref{fig:deltaR}).}

\begin{table}
  \centering
  \begin{tabular}{|l|l|l|} 
      \hline 
      Parameter & Fully resolved & Point particle\\ 
      \hline
      Lattice size $L$ & 96 & 9 \\
      \hline
      Particle radius $a$ & 1.25 & 0.117 \\
      \hline 
      Viscosity $\eta$ & $\frac{1}{6}$ & $\frac{1}{6}$ \\
      \hline
      Density $\rho$ & 1.0 & 1.0 \\
      \hline
      Reynolds number $\mathrm{Re}$ & 0.01 & 0.01 \\
      \hline
      Stokes time $\tau_S$ & 937 & 8 \\
      \hline
      Number of particles & 321 & 321 \\
      \hline
      Volume fraction $\phi$ & $3.0 \times 10^{-3}$ & $3.0 \times 10^{-3}$ \\
      \hline
  \end{tabular}
  \caption{Parameters for simulations used to compare fully-resolved
  and point-like algorithms; see text and figures
  \ref{fig:random-traj-diff} and \ref{fig:deltaR}.}
  \label{tab:params}
\end{table}

\section{Discussion}
The focus of this work has been to {\altered derive and validate a
general method for addressing singular forcing in the LBE, with
specific} application to the simulation of point-like particles. We
have shown the method to agree well with analytic results, where
available, and with fully-resolved particle algorithms at low
concentrations and Reynolds number. Additionally, due to its careful
construction, the regularized $\delta$-function provides a good
interpolation scheme, minimizing velocity fluctuations as the particle
moves relative to the computational grid.  {\altered Indeed we find
that for sedimenting colloids the trajectories are much smoother in
our Stokeslet algorithm than for the fully resolved simulation. In the
latter, the discretization renders particles hydrodynamically
aspherical with shapes that vary as they move across the lattice
{\alteredRWN \cite{Stratford:2005a}}. Absence of such irregularities
in the Stokeslet code may make this generally preferable at small
volume fractions.}

{\altered For dilute suspensions of sedimenting colloids, our new
algorithm can thus perform simulations of accuracy comparable to (or
even better than) that of a fully resolved code, but at vastly reduced
computational cost. As shown in Table~\ref{tab:params}, similar
particle numbers and volume fractions can be simulated with an LB
lattice that is smaller in linear dimension by a factor $\lambda
\approx 10$. (This is the ratio of the particle radii in the two
simulations.) The computational time to update the particle positions
is essentially negligible, so that the CPU time needed to perform one
LB time step is decreased by a factor of $\lambda ^3$; moreover the
Stokes time, $\tau_{S} =\frac{\rho a^2}{\eta \mathrm{Re}}$, scales as
$\lambda^2$. The latter sets the time basic time scale for evolution
of sedimentation trajectories, so that for this problem we expect a
speed-up of $\mathcal{O}(\lambda^5) \approx 10^5$.  This should allow
us to study the sedimentation behaviour of dilute systems with tens of
millions of particles; we hope to pursue this avenue elsewhere.}

\acknowledgments{We thank Kevin Stratford for useful discussions and
K. Vijay Kumar for communications on the Crowley instability. Work
funded in part by EPSRC GR/S10377.}

\appendix
\section{Subtraction procedure}\label{sec:subtraction}

We derive here the correction factor that arises from replacing the
background velocity with the fluid velocity in the Fax\'en
relation. The velocity at a node due a regularized Stokeslet located
at ${\bf R}_s$ is
\begin{equation}
{\bf v}_s({\bf r}, {\bf R}_s) = \frac{1}{L^3}\sum_{\bf k\neq
0}\frac{e^{i\bf k\cdot{\bf r}}}{\eta k^2}\delta^{P}({\bf k};{\bf
R}_s)(\eye - \hat {\bf k} \hat {\bf k})\cdot{\bf F}_s
\end{equation}
This velocity interpolated from the neighbouring nodes to the location
of the particle is
\begin{widetext}
\begin{equation}
\sum_{\bf r}{\bf v}^s({\bf r}, {\bf R}_s)\delta^{P}({\bf r - R}_s) =
{\bf F}_s\cdot\sum_{{\bf r},\bf k\neq 0}\frac{e^{i\bf k\cdot{\bf
r}}}{\eta k^2}\frac{\delta^{P}({\bf k};{\bf R}_s)}{L^3}(\eye - \hat
{\bf k} \hat {\bf k})\delta^{P}({\bf r - R}_s)
\end{equation}
Completing the spatial sum, we get the for the interpolated Stokeslet
velocity,
\begin{equation}
\sum_{\bf r}{\bf v}_s({\bf r}, {\bf R}_s)\delta^{P}({\bf r - R}_s) =
{\bf F}_s\cdot\sum_{\bf k\neq 0}\frac{|\delta^{P}({\bf k};{\bf
R}_s)|^2}{\eta k^2}(\eye - \hat {\bf k}\hat {\bf k}) \equiv \frac{{\bf
F}_s}{6\pi \eta a_L({\bf R}_s)}
\end{equation}
\end{widetext}
which shows that the offset parameter $a_L$ obeys
\begin{equation}
\frac{1}{a_L({\bf R}_s)} = \frac{6\pi}{L^3}\sum_{\bf k\neq
0}\frac{|\delta^{P}({\bf k};{\bf R}_s)|^2}{k^2}(\eye - \hat {\bf
k}\hat {\bf k})
\end{equation}
is indeed independent of viscosity, and particle radius $a$, but
depends on the lattice size $L$ and the numerical implementation of
the regularization and interpolation.

\section{Noise}\label{sec:noise}
Considering a spherical sedimenting particle in the Langevin picture,
we can write
\begin{equation}
  m \ddot{\vec{R}} = - 6\pi\eta a \left( \dot{\vec{R}} -
  \vec{v}^{\infty}\right) + \vec{F} + \vec{\zeta}(t)
\end{equation}
and taking the inertialess limit gives 
\begin{equation}
  \dot{\vec{R}} = \vec{v}^{\infty}(\vec{R}) + \frac{\vec{F}}{6 \pi \eta
    a} + \vec{\zeta}(t)
\end{equation}
We note that in this equation, noise only comes in through the
Gaussian random variable $\vec{\zeta}$ and that in particular the
fluid velocity $\vec{v}^\infty$ is completely deterministic.

The update rule for our model particle is
\begin{equation}
  \dot{\vec{R}} = \vec{v}(\vec{R}) - \frac{\vec{F}}{6 \pi \eta a_L} +
  \frac{\vec{F}}{6 \pi \eta a}
\end{equation}
which is sufficient for the infinite P\'eclet number regime.  If one
uses fluctuating LB, then the interpolated velocity $\vec{v}(\vec{R})$
contains a noise component.  However, we do not expect the magnitude
of the noise to be appropriate for a particle of the desired radius
$a$, since the random component of the velocity has no dependence on
the radius of the particle.  In fact, the variance of this noise is
that expected for a Brownian particle with the same radius as the
offset parameter $a_L$ ($ > a$) and we use this fact to determine its
value from a diffusion ``experiment'' on an unforced particle, as
explained below.

Knowing that the particle will otherwise diffuse as one with a much
larger radius, we add a white noise term to the update rule for the
model
\begin{equation}
  \dot{\vec{R}} = \vec{v}(\vec{R}) - \frac{\vec{F}}{6 \pi \eta a_L} +
  \frac{\vec{F}}{6 \pi \eta a} + \vec{\zeta}^\prime (t)
\end{equation}
The variance of the extra noise is determined, by the requirement of
satisfying the fluctuation-dissipation theorem, to be
\begin{equation}
  \langle \zeta^\prime_i(t) \zeta^\prime_j(t^\prime) \rangle =
  \frac{kT}{3 \pi \eta} \left( \frac{1}{a} - \frac{1}{a_L} \right)
  \delta_{ij} \delta (t - t^\prime).
\end{equation}

To determine the value of the offset parameter $a_L$, we set up a
simulation of a single unforced particle in periodic boundary
conditions at finite temperature and disable the extra noise term
discussed above, giving
\begin{equation}
  \dot{\vec{R}} = \vec{v}(\vec{R})
\end{equation}
as the equation of motion of the particle.

We let the simulation equilibrate for the characteristic time for
momentum to diffuse across the box size, $T ~ L^2 / \eta$, before
recording the displacement as a function of time. This is repeated for
a number of different starting positions relative to the LB grid and a
plot of $\langle \vec{r}^2 \rangle$ vs. $t$ is used to estimate the
diffusivity. We then use the Stokes-Einstein relation to derive a
radius and use this as the offset parameter. We then test to ensure
that this gives the correct sedimentation behaviour of a particle.


\end{document}